\documentstyle[prb,aps,multicol,psfig]{revtex}
                        
\begin{document}

\title{Interatomic Potential for Silicon Defects and Disordered Phases}
                            
\author{Jo\~ao F. Justo$^{(1)}$, Martin Z. Bazant$^{(2)}$, Efthimios
Kaxiras$^{(2)}$, V. V. Bulatov$^{(3)}$, and Sidney Yip$^{(4)}$}

\address{$^{(1)}$Instituto de F\'{\i}sica da Universidade de S\~ao Paulo,
CP 66318, CEP 05315-970 S\~ao Paulo - SP, Brazil}

\address{$^{(2)}$Department of Physics, 
Harvard University, Cambridge, MA 02138}

\address{$^{(3)}$Department of Mechanical Engineering, 
Massachusetts  Institute of Technology, Cambridge, MA 02139}
           
\address{$^{(4)}$Department of Nuclear Engineering, 
Massachusetts  Institute of Technology, Cambridge, MA 02139}
 
\vspace{0.2cm}
            
\date{December 1, 1997}

\maketitle  
\begin{abstract}

	We develop an empirical potential for silicon which represents
a considerable improvement over existing models in describing local
bonding for bulk defects and disordered phases. The model consists of
two- and three-body interactions with theoretically motivated
functional forms that capture chemical and physical trends as
explained in a companion paper. The numerical parameters in the
functional form are obtained by fitting to a set of {\it ab initio}
results from quantum mechanical calculations based on density
functional theory in the local density approximation, which
include various bulk phases and defect structures.  We test the
potential by applying it to the relaxation of point defects, core
properties of partial dislocations and the structure of disordered
phases, none of which are included in the fitting procedure.  For
dislocations, our model makes predictions in excellent agreement with
{\it ab initio} and tight-binding calculations.  It is the only
potential known to describe both the 30$^\circ$- and
90$^\circ$-partial dislocations in the glide set \{111\}.  The
structural and thermodynamic properties of the liquid and amorphous
phases are also in good agreement with experimental and {\it ab
initio} results. Our potential is the first capable of simulating a
quench directly from the liquid to the amorphous phase, and the
resulting amorphous structure is more realistic than with existing
empirical preparation methods.  These advances in transferability come
with no extra computational cost, since force evaluation with our model is
faster than with the popular potential of Stillinger-Weber, thus
allowing reliable atomistic simulations of very large atomic systems.

\end{abstract}
            

%\narrowtext 
\begin{multicols}{2}

\section{Introduction}
 
Silicon is one of the most intensely investigated materials.  Interest
in silicon is mostly motivated by its great technological importance,
but it is also regarded as the prototypical covalent solid, on which
theoretical concepts about covalent bonding can be tested, and new
ideas can be explored.  The nature of covalent bonding often makes the
description of complicated phenomena difficult: the covalent bond can
only be described properly by quantum theory, while many interesting
phenomena involve large numbers of atoms (in the range 10$^3$-10$^9$),
which quantum mechanical approaches cannot handle \cite{payne}.  
In this sense, silicon represents an ideal candidate for modeling by 
an empirical interatomic potential: the development of a potential will 
at once put to a rigorous test our understanding of the physics of 
covalent bonding and, if successful, will enable the simulation of 
important complex processes that involve large number of atoms.

Many attempts to construct an empirical model for interatomic
interactions in Si have already been reported \cite{bala}.  Of all
these models, the Stillinger and Weber (SW) \cite{sw} and the Tersoff
\cite{t1,t2,t3} potentials are the most widely used \cite{bala}.  The
SW potential consists of two- and three-body terms and was fitted to
experimental properties of the diamond cubic (DC) and molten phase of
silicon \cite{sw}.  It has been used, for example, to study lattice
dynamics \cite{li}, point defects \cite{batra,nastar}, surfaces
\cite{poon}, and the liquid and amorphous phases
\cite{sw,broughton,luedtke,kluge,ishi}.  The Tersoff potential (its three
versions usually referred to as T1 \cite{t1}, T2 \cite{t2}, and T3
\cite{t3}) consists of many-body interactions included in a bond order
term and was fitted to {\it ab initio} results for several Si polytypes.  It has
been used to study lattice dynamics \cite{li}, thermomechanical
properties \cite{porter}, point defects \cite{t2,t3}, and the liquid
and amorphous phases \cite{t3,kelires,ishi}.

Although the SW and Tersoff functional forms have enough flexibility
to describe a number of different configurations, their
transferability to a wide class of structures remains in question.
Several models have attempted to improve the description of
configurations far from equilibrium and far from the database used to
construct the potential, by changing the functional form, using higher
order (up to five-body) expansion terms, increasing the number of
fitting parameters (up to 36), or expanding the set of fitted structures
\cite{kp,mff,bolding,tiff}.  Despite the increased complexity,
these models have been able to improve the description of local
configurations only selectively, and often at a considerable increase
in computational cost.  This suggests that a simple extension to more
elaborate functional forms or larger databases does not necessarily
provide better description of covalent bonding.  Considering the lack
of transferability of existing models, it is of interest to develop a
model for silicon with the following ingredients: superior description
of local structures and computationally efficient evaluations of the
energy and interatomic forces.

In this article we present a new empirical potential for silicon using
a theoretically motivated functional form which emphasizes chemical
and physical trends \cite{partI}, and which is determined by fitting
to a fairly small {\it ab initio} database with only 13 parameters.
This potential represents a considerable improvement over existing
models in describing local structures and extended defects.  It
provides a good description of point defects in the bulk, the
concerted exchange path for self-diffusion, and elastic properties of
bulk silicon. It also predicts core structures of partial dislocations
in the glide set \{111\} in excellent agreement with {\it ab initio}
results. Disordered structures and phase transitions are also
well-described, particularly the amorphous phase, which is better
modeled by dynamical simulations using our potential than by any
empirical preparation method.

The article is organized as follows: The functional form is presented
in detail in section \ref{sec:ff}.  The fitting of the model is
discussed in section \ref{sec:fit}, along with tests of
transferability for bulk crystal structures, defects and activation
complexes.  The fitted potential is then used to calculate core
properties of physically relevant dislocations in section
\ref{sec:dis} and the structure of disordered phases in section
\ref{sec:amo}.

\section{Functional Form}
\label{sec:ff}

\begin{figure}[h]
\centering{
\mbox{\psfig{figure=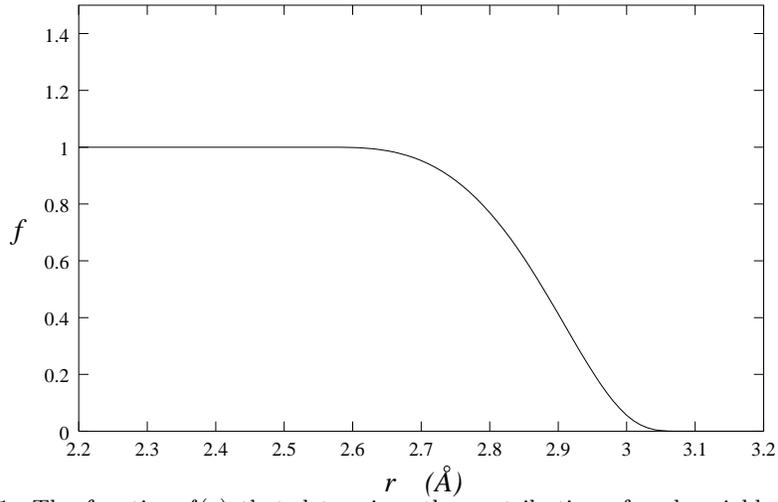,width=4in}}
\begin{minipage}[h]{5in}
\caption{The function $f(r)$ that determines the contribution of each
neighbor to the effective coordination $Z$.}
\label{fig1}
\end{minipage}
}
\end{figure}

	Here we describe the functional form of the new potential,
which we call the Environment Dependent Interatomic Potential (EDIP)
for bulk silicon, and refer the reader to the companion paper
\cite{partI} for theoretical justification of all the terms.  The
energy of a configuration $\{\vec{R}_i\}$ is expressed as a sum over
single-atom energies, $E = \sum_i E_i$, each containing two- and
three-body terms,
\begin{equation}
E_i = \sum_{j\neq i} V_2(R_{ij}, Z_i) + \sum_{j\neq i} \sum_{k \neq i,
 k>j} V_3(\vec{R}_{ij},\vec{R}_{ik}, Z_i) ,
\end{equation}
where $ V_2(R_{ij}, Z_i)$ is an interaction between atoms $i$ and $j$
representing pairwise bonds, and $ V_3(\vec{R}_{ij},\vec{R}_{ik},
Z_i)$ is the interaction between atoms $i$, $j$ and $k$ centered at
atom $i$ representing angular forces. Both types of interactions
depend on the local environment of atom $i$ through its effective
coordination number, defined by,
\begin{equation}
Z_i = \sum_{m \neq i} f(R_{im})
\end{equation}
where $f(R_{im})$ is a cutoff function that measures the contribution
of neighbor $m$ to the coordination of atom $i$ in terms of the
separation $R_{im}$.  The neighbor function, depicted in
Fig. \ref{fig1}, is unity for $r<c$, with a gentle drop from 1 to 0
between $c$ and $a$, and is 0 for $r>a$:
\begin{equation}
f(r) = \left\{ \begin{array}{ll} 1 & \mbox{if $r < c$} \\
               \exp\left(\frac{\alpha}{1 - x^{-3}}\right) & \mbox{if
               $c < r < a$} \\ 0 & \mbox{if $r > a$} \end{array}
               \right.
\end{equation}
where $x = \frac{(r- c)}{(a - c)}$.  A neighbor of atom $i$ at a
distance $r<c$ is considered a full neighbor, while the neighbors
between $c$ and $a$ give only a partial contribution to $Z_i$.  The
cutoffs are constrained to reproduce the coordinations of important
crystal structures, {\it e. g.} $Z_i = 4$ for the diamond lattice.

\begin{figure}[h]
\begin{center}
\leavevmode
\mbox{
\psfig{figure=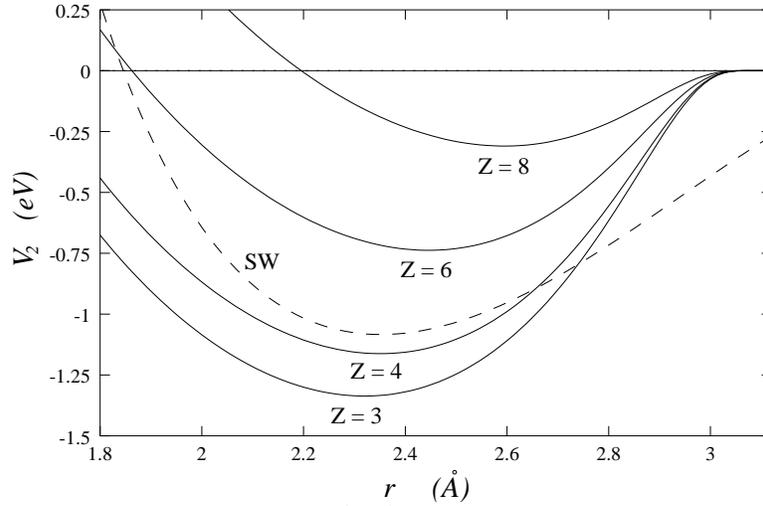,width=4in}
}
\begin{minipage}[h]{5in}
\caption{The two-body interaction $V_2(r,Z)$ as a function of
separation $r$ shown for coordinations 3, 4, 6 and 8 and compared with
the Stillinger-Weber (SW) pair interaction.}
\label{fig2}
\end{minipage}
\end{center}
\end{figure}

	The two-body term includes repulsive and attractive interactions,
\begin{equation}
V_2(r,Z) = A \left[ \left(\frac{B}{r}\right)^\rho - p(Z) \right] \exp
\left( \frac{\sigma}{r-a} \right) ,
\end{equation}
which go to zero at the cutoff $r=a$ with all derivatives continuous.
Although this choice is quite similar to the SW two-body term (and indeed
reduces to it exactly for small distortions of the diamond lattice),
the bond strength adapts to changes in the local atomic environment.
The coordination dependence introduces an asymmetry, $V_2 (R_{ij},Z_i)
\neq V_2 (R_{ji},Z_j)$, similar to the Tersoff potential in that the
strength of the attractive force is controlled by a bond order
function $p(Z)$ that depends on the local coordination.  This
dependence is motivated by theoretical calculations which have
demonstrated the weakening of the attractive interaction and the
corresponding increase in bond length for increasing coordination
\cite{partI,cfm,bazant}.  These studies reveal a shoulder in
the function $p(Z)$ at the ideal coordination $Z_0=4$, where the
transition from covalent ($Z\leq Z_0$) to metallic ($Z>Z_0$) bonding
occurs.  This theoretical dependence can be accurately captured by a
Gaussian function \cite{partI},
\begin{equation}
p(Z) = \mbox{e}^{-\beta Z^2} \ .
\end{equation}
Figure \ref{fig2} shows the $V_2 (r,Z)$ term for several
coordinations, compared to the two-body part of 
the SW potential $V_2^{SW} (r)$.
Note that the fitted $V_2 (r,Z)$ resembles closely the inverted {\it
ab initio} pair potentials for silicon crystals with the same
coordinations \cite{bazant}, a feature built into the function form
\cite{partI}. 

	The three-body term contains radial and angular factors,
\begin{equation}
V_3(\vec{R}_{ij},\vec{R}_{ik}, Z_i) = g(R_{ij}) g(R_{ik})
h(l_{ijk},Z_i) ,
\end{equation}
where $l_{ijk} = \cos \theta_{ijk} =
\vec{R}_{ij}\cdot\vec{R}_{ik}/R_{ij}R_{ik}$.  The radial function is
chosen to have the SW form,
\begin{equation}
g(r) = \exp \left( \frac{\gamma}{r-a}\right) ,
\end{equation}
and goes to zero smoothly at the cutoff distance $a$.  The angular
function $h(l,Z)$ is strongly dependent on the local coordination
through two functions $\tau(Z)$ and $w(Z)$ that control the
equilibrium angle and the strength of the interaction, respectively.
Theoretical considerations lead us to postulate the following form
\cite{partI}:
\begin{equation}
h(l,Z) = H\left( \frac{ l + \tau(Z) }{w(Z)} \right) ,
\end{equation}
where $H(x)$ is a generic function satisfying the constraints,
$H(x)>0$, $H(0)=0$, $H'(0)=0$ and $H"(0)>0$.  Our specific choice is
\begin{equation}
h(l,Z) = \lambda 
\left[ \left( 1 - e^{ - Q(Z) (l + \tau(Z))^2} \right) +
\eta Q(Z) (l + \tau(Z))^2 
\right] .
\end{equation}
We make the choice $w(Z)^{-2} = Q(Z) = Q_0 e^{-\mu Z}$ to control the
strength of angular forces as function of coordination.  The
three-body angular function becomes flatter (and hence angular forces
become weaker) as coordination increases, representing the transition
from covalent bonding to metallic bonding.  The angular function
$h(l,Z)$ has two major contributions. The first, $H_1(x) \propto 1 -
\mbox{e}^{-x^2}$, is symmetric about the minimum, becoming flat for
small angles. This choice of shape for the angular function was also
used by Mistriotis, Flyntzanis and Farantos (MFF) \cite{mff}, but due to its
environment-dependence our angular function is fundamentally
different. In a preliminary fitted version of EDIP, we included only
this flat, symmetric term \cite{mrs}, but we found that it is not
suitable for several structures (particularly the amorphous
phase). Indeed, a more asymmetrical angular function is suggested by
approximations of quantum-mechanical (tight-binding) models
\cite{alinaghian,carls,carlsson} and exact inversions of {\it ab
initio} cohesive energy curves \cite{bazant}, so the present form of
EDIP also contains a second term, $H_2(x) \propto x^2$ (identical in
shape to the SW form but containing environment-dependence), 
which gives a stronger interaction for small angles.

\begin{figure}[h]
\centering{
\mbox{\psfig{figure=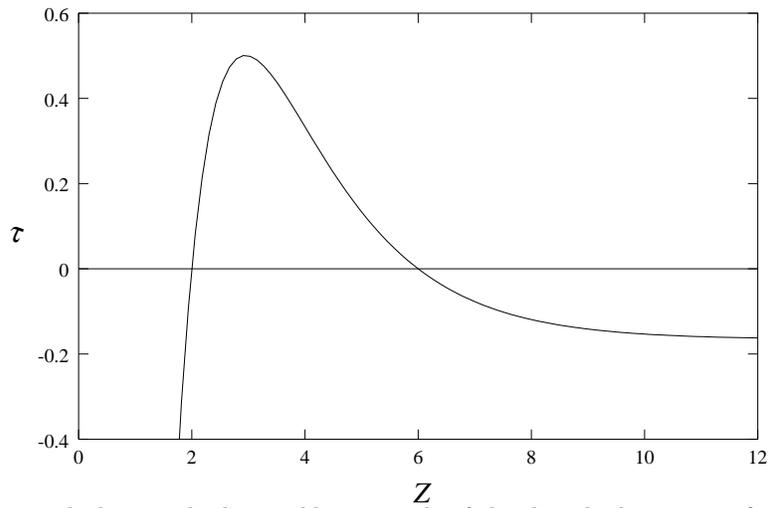,width=4in}}
\begin{minipage}[h]{5in}
\caption{$\tau$, which controls the equilibrium angle of the
three-body term, as function of coordination.}
\label{fig2a}
\end{minipage}
}
\end{figure}

\begin{figure}[h]
\begin{center}
\leavevmode
\mbox{
\psfig{figure=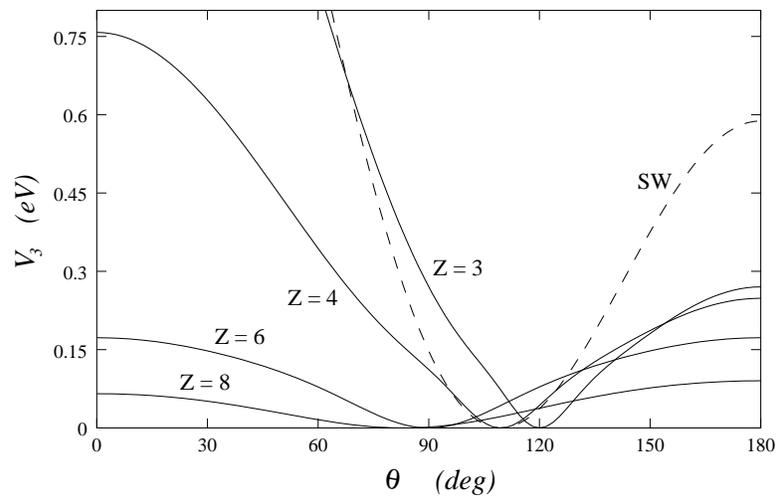,width=4in}
}
\begin{minipage}[h]{5in}
\caption{The three-body interaction $V_3(r,r,\cos\theta,Z)$ for a pair
of bonds of fixed length $r = 2.35$ {\AA}\ subtending an angle
$\theta$, shown for coordinations 3, 4, 6 and 8 and compared with the
Stillinger-Weber three-body interaction.}
\label{fig3}
\end{minipage}
\end{center}
\end{figure}

The function $\tau (Z) = - l_0(Z) = - \cos({\theta_0(Z)})$ controls
the equilibrium angle $\theta_0(Z)$ of the three-body interaction as a
function of coordination.  This feature of the potential makes it
possible to model the proper hybridization of atoms in different
environments \cite{partI,beigi}: If a silicon atom is three- or
four-fold coordinated, it will prefer to form $sp^2$ or $sp^3$ hybrid
bonds with equilibrium angles $\theta_0(3) = 120^\circ$ and
$\theta_0(4)=109.471^\circ$, respectively.  For coordinations 2 and 6,
we take $\tau(2) = \tau(6) = 0$ or $\theta_0(2) = \theta_0(6) =
90^\circ$. For two-fold coordination, this choice describes the
preference for bonding along two orthogonal $p$-states with the
low-energy nonbonding $s$ state fully occupied. For six-fold
coordination, the choice $\theta_0(6) = 90^\circ$ reflects the $p$
character of the bonds (which are also metallic, as discussed below).
We construct $\tau (Z)$ to interpolate smoothly between the special
points ($Z = 2, 3, 4, 6$) with the following form,
\begin{equation}
\tau (Z) = u_1 + u_2 (u_3 e^{-u_4 Z} - e^{-2 u_4 Z}) ,
\end{equation}
with the parameters chosen as $ u_1 = -0.165799$ , $u_2 = 32.557$,
$u_3 = 0.286198$, and $u_4 = 0.66$, resulting in the curve shown in
Fig. \ref{fig2a}, which is consistent with the results of
quantum-mechanical approximations \cite{pZnote}. Note that these
parameters are theoretically determined and are not allowed to vary in
the fitting described in the next section.  Figure \ref{fig3} shows
the three-body term $V_3(\vec{R}_{ij},\vec{R}_{ik}, Z_i)$ for three
atoms at a distance $2.35$ \AA, and for different coordinations.  We
also compare the three-body term to the SW three-body potential
$V_3^{SW} (\vec{R}_{ij},\vec{R}_{ik})$.  The SW angular form $h_{SW}
(\theta_{jik}) = \lambda_{SW} (\cos(\theta_{jik}) + 1/3)^2$ penalizes
the configurations with angles smaller than $90^\circ$ with a large
positive contribution.  In contrast, the angular function of our model
potential gives a considerably weaker interaction at small angles.

In summary, this implementation of EDIP for bulk silicon has 13
adjustable parameters: $A$, $B$, $\rho$, $\beta$, $\sigma$, $a$, $c$,
$\lambda$, $\eta$, $\gamma$, $Q_0$, $\mu$ and $\alpha$.  It also has
continuous first and second derivatives with respect to the atomic
position vectors. The functional form already contains considerable
information about chemical bonding in bulk silicon taken directly from
theoretical studies, mostly of ideal crystal structures. The
(relatively few) adjustable parameters provide the necessary
freedom to extrapolate these bonding dependences for defect structures
strictly outside the theoretical input, as described below.

We close this section by addressing the crucial issue of computational
efficiency (which motivates the use of empirical methods in the first
place).  The environment terms in the two- and three-body iterations
require extra loops in the force calculation.  In the case of the
three-body interaction, it requires a four-body loop, that would make
a force evaluation more expensive than with the SW potential.
However, the four-body loop needs to be performed only for those
neighbors $l$ of atom $i$ in which $\partial f (R_{il})/ \partial
R_{il} \neq 0$. This happens only when atoms are in the range $c<r<a$,
i.e. only for a small number of the neighbors.  Therefore, one force
evaluation using our model is approximately as efficient as one using
the SW potential, showing that increased sophistication and realism
can be achieved with insignificant computational overhead.  In fact,
since the fitted cutoff distance (given in the next section) is
smaller than the corresponding SW cutoff, computing forces with our model
is typically faster than with the SW potential. For example, in liquid
phase simulations on a Silicon Graphics R-10000 processor it takes 30
$\mu$sec/atom to evaluate forces using EDIP, compared with 50
$\mu$sec/atom using the SW potential.

\section{Fitting and Tests}
\label{sec:fit}

In order to determine the values of the adjustable parameters, we fit
to a database that includes {\it ab initio} results \cite{kaxlda}, 
based on density functional theory in the 
local density approximation (DFT/LDA), for 
bulk properties (cohesive energy and lattice constant of the DC
structure), selected values along the unrelaxed concerted exchange
(CE) path \cite{pand} for self-diffusion, some formation energies of
unrelaxed point defects (vacancy and interstitial at the tetrahedral
and hexagonal configurations) \cite{baryam,kelly,seong}, a few key
values in the generalized stacking fault (GSF) energy surface
\cite{kax1}, and the experimental elastic constants \cite{simm}.  This
rich set of configurations contains many of the important local
structures found in condensed phases and bulk defects and thus
improves the transferability of the model.  The database does not
include other high symmetry configurations, such as SC, BCC or FCC (as
many of the existing empirical potentials have done), although the DC
structure is required to be the lowest in energy. These hypothetical
metallic structures have large enough energies compared to covalent
ones to be considered irrelevant.  Incidentally, fitting the model
simultaneously to such a wide set of configurations and properties
represents a considerable computational challenge.  The fitting is
accomplished using a least-squares approach, with each configuration
in the database weighted appropriately.  All parameter values are
allowed to vary at once through a simulated annealing process.
Theoretical estimates of the parameters were also used to restrict the
range of parameter-space that needed to be explored \cite{partI}.
Table \ref{tab1} provides the best set of parameters we obtained.

	The calculated energies and properties of several bulk
structures as obtained from our potential and from {\it ab initio} and
other empirical potential calculations are given in Table \ref{tab2}.
This compilation of results includes the DC structure which is the
ground state of Si, and several other high-coordination bulk
structures.  Although the latter structures were not included in the
fitting database, our model provides a reasonable description of their
energies and a good description of equilibrium lattice constants.
Structures such as $\beta$-tin and BCT5\cite{kaxboyer} are also of
interest because they have low-energy and relatively low-coordination
(5 for BCT5 and 6 for $\beta$-tin) \cite{bala}.
Experimentally, the DC phase transforms into the $\beta$-tin structure
under pressure.  For the $\beta$-tin structure, our model predicts a
cohesive energy per atom higher than that of DC by $\Delta E = 0.67$
eV, and a lattice parameter $a_o = 4.76 $ {\AA}, as compared to {\it
ab initio} results of 0.21 eV and 4.73 {\AA} respectively.  For the
BCT5 structure, our model predicts $\Delta E = 0.26$ eV and $a_o =
3.36$ {\AA}, as compared to {\it ab initio} results of 0.23 eV and
3.32 {\AA}, respectively.

	Most existing empirical potentials give a poor (or marginally
acceptable) description of elastic properties of the DC crystal, which
directly affects the description of the crystal deformation. In the
fitting database we included the three independent elastic constants
$C_{11}$, $C_{12}$ and $C_{44}$ from experiment.  Table \ref{tab3}
compares elastic constants, obtained with the homogeneous deformation
method \cite{wallace}, as given by our model with the results from
other empirical potentials and with experimental results.  The shear
constant $C_{44}$ is particularly important for the description of
long range interactions \cite{bala}. Although $C_{44}$ is
underestimated by most empirical models, ours is reasonably close to
the experimental value. Our potential also almost perfectly reproduces
the {\it ab initio} value of the shear modulus $C_{44}^\circ$ without
internal relaxation\cite{nielsen}, as anticipated by theory
\cite{partI}.  Table \ref{tab3} also includes other elastic
properties, such as the second shear constant, $C_{11} - C_{12}$, and
the Cauchy discrepancy, $\Delta C = C_{12} - C_{44}$, both important
for determining crystal stability \cite{jinghan}.  By predicting a
negative Cauchy discrepancy, our potential offers a qualitative
improvement over most other existing potentials \cite{bala} and
several tight-binding (TB) models \cite{noam} which give positive
values.  It also provides a quantitative improvement over other TB
models that get the correct sign of $\Delta C$ \cite{kwon94}. In
summary, our potential gives elastic constants in excellent agreement
with experimental and {\it ab initio} results. As we have shown in
Ref. \cite{partI}, accurate elastic behavior is not simply the result
of good fitting, but is rather a built-in feature of our functional
form.

	Point defects in the DC crystal involve large atomic
relaxations and rebonding, thus representing the first test for the
transferability of our model in describing local structures away from
equilibrium. Our fitting database included the unrelaxed structures of
the vacancy (V) and the interstitial in tetrahedral (I$_T$) and
hexagonal (I$_H$) configurations \cite{kaxlda}.  Table \ref{tab4}
shows the formation energy for the unrelaxed and relaxed structures of
V, I$_T$ and I$_H$ as obtained using our model, compared to {\it ab
initio} \cite{baryam,kelly,seong} and TB calculations \cite{noam}, and
from calculations \cite{bala} using the SW and Tersoff (T2 and T3)
potentials.  The relaxed structures are computed using an energy
minimization conjugate-gradient method \cite{vas1}.  Although the SW
and Tersoff potentials give a reasonable description of relaxed
structures, they clearly fail in describing the energy released upon
relaxation. In our model, on the other hand, the relaxation energies
are much lower and in reasonably good agreement with {\it ab initio}
calculations.  We emphasize that none of the relaxation energies or
structures were used in the fitting database. The $<110>$ split
interstitial (the lowest energy interstitial configuration) is well
described by our model, in spite of also not being included in the
fitting.  The formation energy for the relaxed structure of the
$<110>$ split interstitial is 3.35 eV in our model, compared to 3.30
eV from {\it ab initio} calculations\cite{blochl}, while the SW
potential gives 4.68 eV.

\begin{figure}[h]
\centering{
\mbox{\psfig{figure=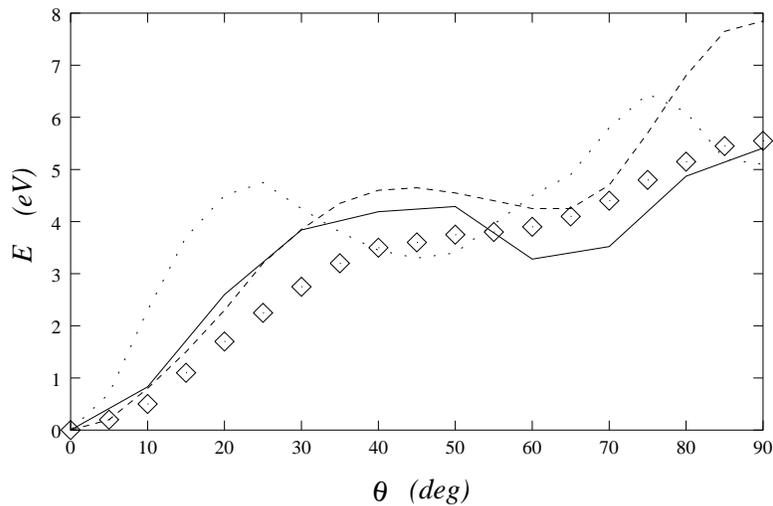,width=4in}}
\begin{minipage}[h]{5in}
\caption{Energy of the concerted exchange path obtained from
calculations using DFT/LDA \protect{\cite{kaxlda,pand}} (diamonds),
the SW (dashed line), the Tersoff (dotted line), and the present
interatomic potential (solid line). The results from {\it ab initio},
SW and Tersoff are from Ref. \protect{\cite{kp}}.}
\label{fig4}
\end{minipage}
}
\end{figure}

	The CE process \cite{pand} has been identified as a possible
mechanism for self-diffusion in Si \cite{kaxpand}.  Most of the
empirical potentials provide only a fair description of this important
and complicated sequence of configurations \cite{kp}.  Fig.
\ref{fig4} shows the energy for the unrelaxed CE path \cite{kp} from
calculations based on DFT/LDA \cite{kaxlda}, the present model and the
SW and Tersoff  potentials.  The results using our model agree
reasonably well with those from DFT/LDA calculations, and are
considerably better than those using the SW or Tersoff potentials.
The activation energy for CE (an important quantity that enters in the
calculation of diffusion rates)
%
% TIM: isn't the relaxed activation energy 4.82eV more relevant?
%
obtained from our model is 5.41 eV, compared to 5.47 eV from {\it ab
initio} calculations, 7.90 eV from the SW potential, and 6.50 eV from
the Tersoff potential. The energy of the relaxed structure, which was
not included in the fitting database, is 4.82 eV using our model, in
good agreement with the {\it ab initio} result of 4.60 eV.

	The fitting database also included selected unrelaxed
configurations of the generalized stacking fault (GSF) energies.  We
considered three points for the the glide set and three for the
shuffle set \cite{kax1} of the \{111\} glide plane.  Table \ref{tab4a}
shows the unstable stacking fault energy for the unrelaxed glide and
shuffle $\{111 \}$ planes as obtained from calculations using DFT/LDA
\cite{kax1}, SW, and the present interatomic potential. Our model
gives good agreement with DFT/LDA calculations for the $<112>$ glide
set, but underestimates the energy for the $<110>$ glide set.  Since
our potential is short-ranged, it gives zero energy for the stable
stacking fault, while the experimental value is 0.006 eV/{\AA}$^2$.
The SW and Tersoff models, which are also short-ranged, give the same
zero energy for the stable stacking fault. The TB model \cite{hansen},
on the other hand, gives 0.005 eV/{\AA}$^2$, in agreement with DFT/LDA
results.  Nevertheless, given that the accuracy of empirical models is
rarely better than a few tenths of an eV per atom, our vanishing
stable stacking fault energy is not particularly problematic.

	In summary, our potential provides an excellent description of
configurations near equilibrium as well as a wide range of point
defects in the DC structure. Although a number of these properties
were explicitly fit, the superior description has been achieved with a
small number of adjustable parameters that is greatly exceeded by the
number of degrees of freedom inherent in these configurations. Thus,
we suggest that it is our physically appropriate functional form
rather than a flexible fitting strategy that has led to the improved
description, a conclusion that is further supported by our results for
extended defects and disordered phases discussed next.

\section{Dislocations}
\label{sec:dis}

	A stringent test of the transferability of our model to local
structures is obtained by calculating core properties of dislocations,
about which no information was included in the fitting database.
Several {\it ab initio} \cite{bigg,huang,arias} and tight-binding
calculations \cite{hansen,nunes} have been performed for dislocations
in silicon.  Although such calculations are feasible only for small
systems containing of order few hundred atoms, they provide important
information, such as the core structure
of the 90$^\circ$-partial dislocation \cite{bigg}, the kink structure in
the 30$^\circ$-partial dislocation \cite{huang}, and dislocation
interactions \cite{arias}.  However, full simulations of dislocation
dynamics and its effect on the mechanical properties of the solid 
require much larger cells owing to the long range
interaction of the stress fields, and therefore can only be performed
using methods that are computationally less expensive.  Empirical
models have been used to study several aspects of dislocations in
silicon \cite{vas1,trinc,nande}.  No such model has proven reliable
enough to provide a description of long range \cite{trinc} as well as
core properties \cite{duesb1} of dislocations at the same time.  The
flaw in describing long range interactions is due to a poor
description of elastic forces \cite{trinc,teichl}, while the
flaw in describing core properties is due to a poor description of
local structures \cite{duesb1}.  Existing models do not give the
correct structure for reconstructed cores: for example, the SW
potential gives reconstruction only for the 30$^\circ$-partial
dislocation, and the Tersoff potential gives reconstruction only for
the 90$^\circ$-partial.

\begin{figure}[h]
\centering{
\mbox{\psfig{figure=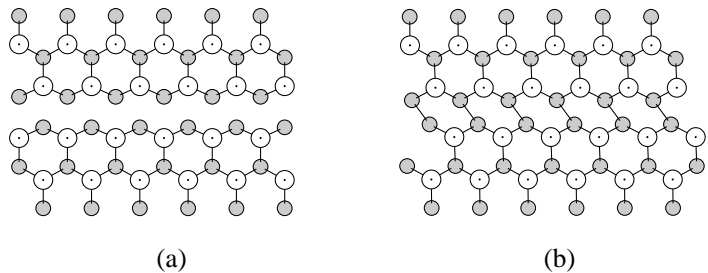,width=4in}}
\begin{minipage}[h]{5in}
\caption{Atomic structure in the core of a $90^\circ$-partial dislocation
on the (111) plane. (a) Symmetric and (b) asymmetric reconstructions. Open
and shaded circles represent atoms in two different (111) planes. }
\label{fig6}
\end{minipage}
}
\end{figure}

	In the present study, all dislocation structures are computed
using energy minimization methods \cite{vas1} at constant stress
\cite{pr} for a system of 3600 atoms, and the cell boundaries lie in
the $[11\overline{2}]$, $[111]$, and $[\overline{1}10]$ directions.
Fig. \ref{fig6} shows (a) the unreconstructed and (b) the
reconstructed core structure of a $90^\circ$-partial dislocation. The
reconstruction energy is defined as the energy gain (per unit length)
for the system to go from unreconstructed to reconstructed
configuration.  Table \ref{tab5} compares the reconstruction energy in
units of eV/$b$ (where $b$ is the unit length of the dislocation, $b =
3.84${\AA}), as given by our model and by SW, Tersoff, TB, and {\it ab
initio} calculations.  Configuration (b) is neither stable nor
metastable for the SW potential, i.e. the SW model does not support
reconstruction for the 90$^\circ$-partial dislocation \cite{duesb1}.  The
present model predicts configuration (b) as the lowest in energy,
while configuration (a) is metastable. This model gives reconstruction
energy of 0.84 eV/$b$, in excellent agreement with the {\it ab initio}
value of 0.87 eV/$b$ \cite{bigg}.  In our calculation, the
reconstructed bonds are stretched by 2.1 \%, while {\it ab initio}
\cite{bigg} and TB \cite{nunes} calculations give bonds stretched by
2.6 \% and 3.0 \% respectively.

\begin{figure}
\centering{
\mbox{\psfig{figure=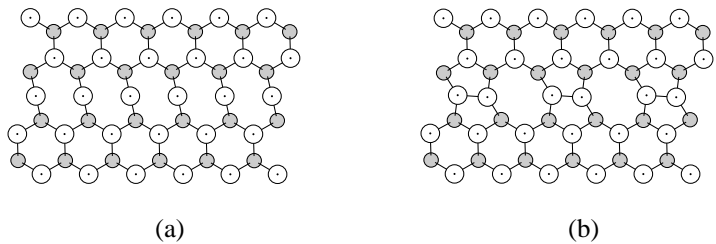,width=4in}}
\begin{minipage}[h]{5in}
\caption{Same as Fig. \ref{fig6} for a $30^\circ$-partial dislocation. 
(a) Unreconstructed and (b) reconstructed configurations.}
\label{fig7}
\end{minipage}
}
\end{figure}

	Fig. \ref{fig7} shows (a) the unreconstructed and (b) the
reconstructed core structures of a 30$^\circ$-partial dislocation.  The
results for the reconstruction energy from {\it ab initio}
calculations \cite{arias2}, and from empirical potential calculations
using the SW, Tersoff and our model, are presented in table
\ref{tab5}.  The Tersoff potential gives negative reconstruction
energy, so that the unreconstructed configuration (a) would be the
more stable \cite{duesb1}, contrary to experimental and {\it ab
initio} results.  Although the SW model gives the correct
reconstructed configuration, the reconstruction energy is twice as
large as the {\it ab initio} result.  Our model, on the other hand,
gives the reconstruction energy in very good agreement with the {\it
ab initio} calculation.  We find that the reconstructed bonds in the
dislocation core are stretched by $3.6 \%$.  The {\it ab initio}
calculations \cite{arias2} give bond stretching of $4.2 \%$ at most.

	It is important to mention that we found a few metastable
partially reconstructed configurations between the unreconstructed and
reconstructed configurations for both the 90$^\circ$- and 30$^\circ$-partial
dislocations.  Such configurations are artifacts of our description of
the local environment during changes in coordination, and probably
have no physical meaning \cite{vas1}.

\begin{figure}
\centering{
\mbox{\psfig{figure=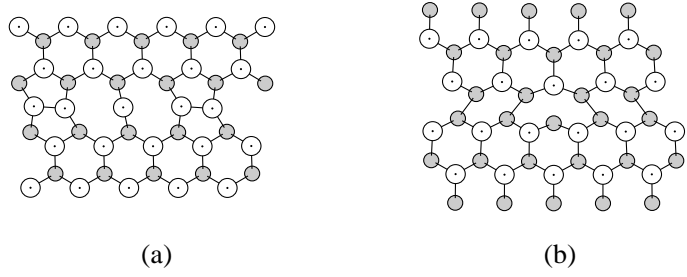,width=4in}}
\begin{minipage}[h]{5in}
\caption{Anti-phase defects (APD) for (a) the  $30^\circ$-partial  
and (b) the $90^\circ$-partial dislocations. }
\label{fig8}
\end{minipage}
}
\end{figure}

	A defect in the core of a reconstructed dislocation is called
an antiphase defect (APD)\cite{suzuki}.  Fig. \ref{fig8} shows APD
configurations in (a) a $30^\circ$-partial dislocation and (b) a
$90^\circ$-partial dislocation.  Table \ref{tab5} gives the corresponding
APD formation energies.  Since the SW model does not produce a
reconstructed configuration for 90$^\circ$-partial dislocations, there is
no stable APD configuration for this case.  For a $30^\circ$-partial
dislocation, the SW model gives an APD formation energy much larger
than {\it ab initio} calculations \cite{arias2}.  The Tersoff
potential gives a negative value for the APD formation energy of the
$30^\circ$-partial and a considerably smaller energy for the APD in a
$90^\circ$-partial.  Our model gives the APD formation energy in good
agreement with {\it ab initio} calculations for the 30$^\circ$-partial
dislocation.  For the $90^\circ$-partial dislocation, to our knowledge
there is no {\it ab initio} APD energy calculation available, but our
APD energy is somewhat low compared to TB calculations \cite{nunes}.

	From the above comparisons we have established that our
empirical model is the first to give a full description of core
properties of both the $90^\circ$ and $30^\circ$-partial dislocations in
silicon with reasonable accuracy.  Our model predicts reasonable 
reconstruction energies and provides a good description of local
deformations in the dislocation cores. These results demonstrate 
remarkable transferability since the local atomic configurations
present in dislocation cores are quite different from the structures
included in the fitting database.

\section{Disordered Structures}
\label{sec:amo}

	Another stringent test of the transferability of our model for
bulk material is the calculation of structural properties of the
liquid and amorphous phases. Such disordered structures contain a rich
set of local bonding states from covalent (amorphous) to metallic
(liquid) about which no information was included in our fitting
procedure.  Existing {\it environment-independent} potentials have had
considerable difficulty in simultaneously describing the crystalline,
liquid and amorphous phases \cite{bala,ishi,stichliq,stichamo,ding}.
In the preceding sections we have demonstrated an improved description
of the crystalline solid and its defects, so we now turn our attention
to whether our environment-dependence can extrapolate these successes
to the liquid and amorphous phases.

	{\it Liquid phase.}  The SW potential has been shown to
reproduce the pair correlation function $g(r)$ of the liquid
\cite{sw,ishi}. It also predicts the melting temperature $T_m$ to
within a few hundred degrees of the experimental value of 1685 K
\cite{bala} (although it was explicitly fit to reproduce $T_m$
\cite{sw}).  In spite of these successes, the SW potential has
difficulty in describing the structure of liquid \cite{ishi}, i.e. it
does not reproduce the {\it ab initio} bond-angle distribution, overly
favoring angles near tetrahedral \cite{stichliq}.  The Tersoff
potentials, on the other hand, predict a $g(r)$ that favors the
unphysical four-fold coordination \cite{t2,t3}, and the only version
predicting reasonable liquid structure is T3 \cite{t3,ishi}.  The
melting temperature, however, is greatly overestimated by T3 at around
3000 K \cite{t3,ishi}. The first-neighbor bond angles and coordination
statistics are not well described by T3, although the statistics are
improved by using an (arbitrarily) longer coordination cutoff beyond
the first minimum of $g(r)$ \cite{ishi}. We emphasize that the T2
potential, the most successful parameterization of the Tersoff model
overall \cite{bala}, cannot decribe the liquid phase \cite{t2}.

\begin{figure}[h]
\begin{center}
\leavevmode
\mbox{
\psfig{figure=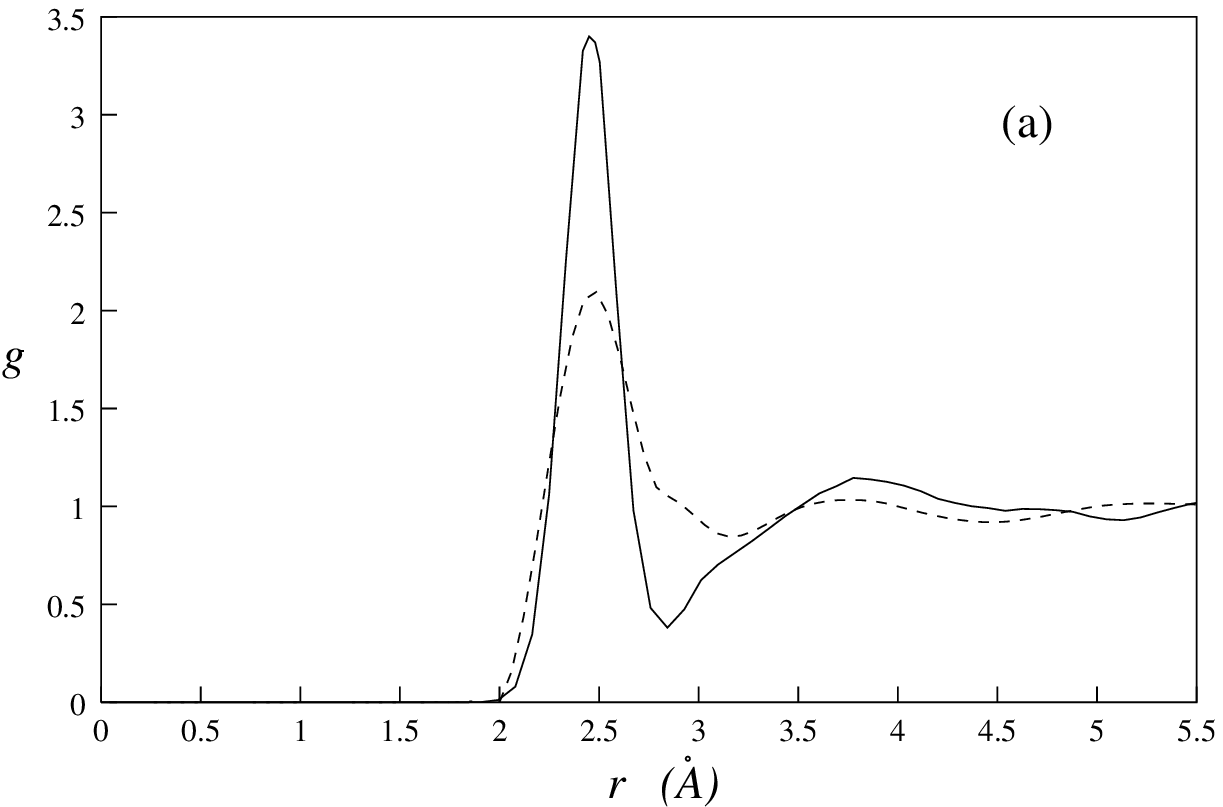,width=4in}
}
\mbox{
\psfig{figure=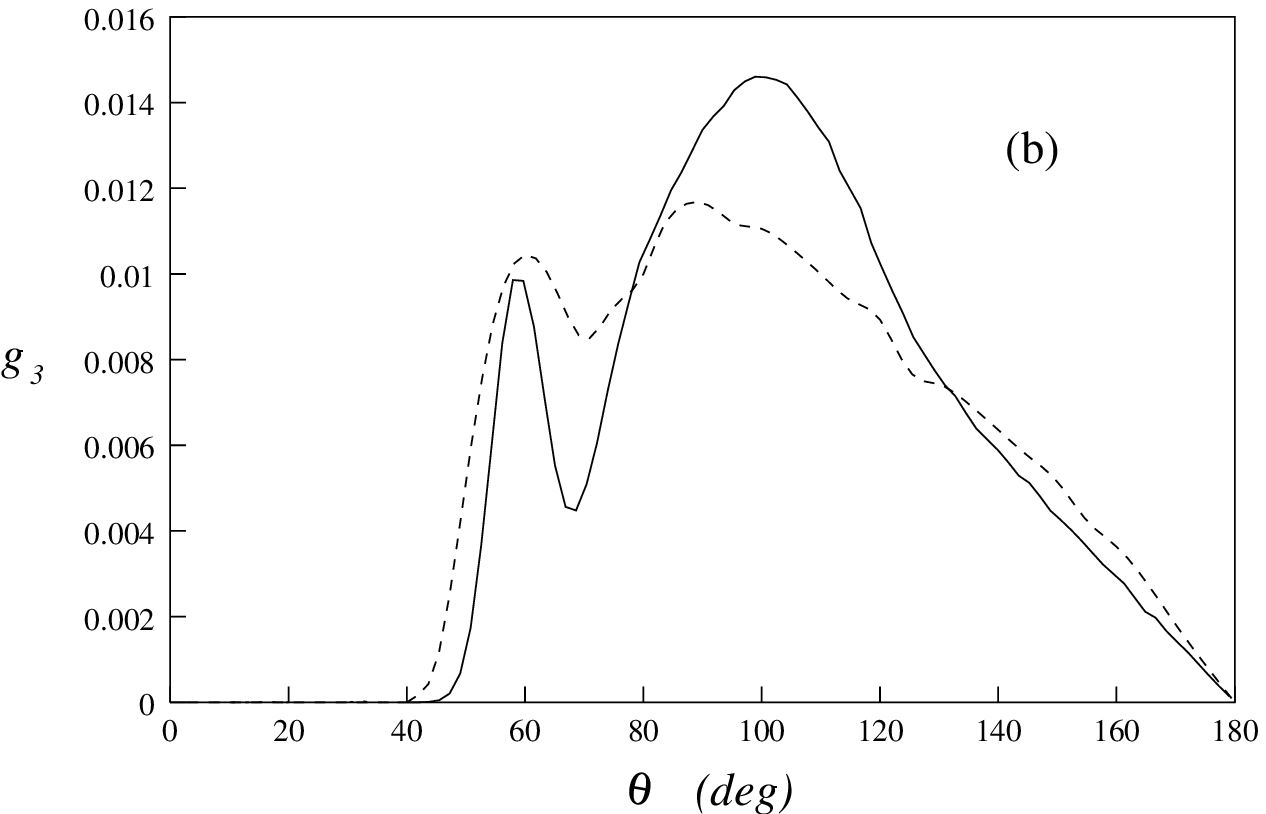,width=4in}
}
\begin{minipage}[h]{5in}
\caption{(a) Pair correlation function and (b) bond angle distribution
for the liquid at $T = 1800$ K and zero pressure with the present
potential (solid lines) and the {\it ab initio} model of
Ref. \protect{\cite{stichliq}} (dashed lines).}
\label{figliq}
\end{minipage}
\end{center}
\end{figure}

	We prepared a 1728-atom liquid sample with the present
potential at $T = 1800$ K and zero pressure using standard simulation
techniques, although we used a considerably longer time and larger
system size than in previous studies \cite{liqprep,allen}. The structural
properties of the model liquid are shown in Fig. \ref{figliq} and
compared with the results of a 64-atom {\it ab initio} molecular
dynamics study \cite{stichliq} (which are similiar to recent results
with 343 atoms including electron spin effects and gradient
corrections \cite{stichliq2}). The pair correlation function $g(r)$
shows excessive short-range order with our potential, as evidenced by
the overly sharp first neighbor peak containing around 4.5 first
neighbors, smaller than the experimental value of 6.4.  This is
consistent with the fact that in our model the density of the liquid
is somewhat smaller than that of the solid, while in reality silicon
becomes 10\% more dense upon melting \cite{glazov}. Although these
features are unphysical, the present model offers a qualitative
improvement in the bond angle distribution function $g_3(\theta,r_m)$,
which gives the (normalized) distribution of angles between pairs of
bonds shorter than $r_m$, the first minimum of $g(r)$. As shown in
Fig.  \ref{figliq}, our potential predicts the auxiliary maximum at
$\theta = 60^\circ$, although the primary maximum is shifted toward
the tetrahedral angle away from the {\it ab initio} most probable
angle of $\theta = 90^\circ$. The present model is the first to
capture the bimodal shape of the first-neighbor bond angle
distribution \cite{ishinote}.

	The thermodynamic properties of the melting transition (aside
from the change in density) are reasonably well-described by our
model, in spite of its not having been fit to
reproduce any such quantities.  The bulk melting point $T_m$ predicted
by our model is 1370 $\pm$ 20 K, which is 20\% below
the experimental value. The melting point was measured for a finite sample
with $(100)2\times 1$ surfaces, heated from 300 K to 1500 K in 2 ns
(over 10 million time steps). A bulk solid with periodic boundary conditions
superheats and melts around 2200 K. The latent heat
of melting is 37.8 kJ/mol, in reasonable agrement with the
experimental value of 50.7 kJ/mol, closer than the SW value of 31.4
kJ/mol \cite{luedtke}.

In summary, although the liquid has some unphysical features with our
potential, it offers some improvements, particularly in describing
bond angles. It is important to emphasize that reasonable liquid
properties are predicted by our model without any explicit fitting to
the liquid phase; in contrast, the only two potentials reported to
give an adequate description of the liquid, SW and MFF, were each fit
to reproduce the melting point \cite{sw,mff}. With our model, the
reduced density and excess of covalent bonds may be artifacts of the
short cutoff of our potential, which is appropriate for the
covalently-bonded structures used in the fitting, but is perhaps too
short to reproduce overcoordination in metallic phases like the
liquid\cite{cutoff_note}.

	{\it Amorphous phase.} Experimentally, amorphous silicon is
known to form a random tetrahedral network, with long-range disorder
and short-range order similar to that of the crystal
\cite{fortner,kugler}. {\it Ab initio} molecular dynamics simulations
of quenching a 64-atom liquid predict almost 97\% four-fold
coordination \cite{stichamo}. An empirical potential would be
invaluable in exploring larger system sizes and longer relaxation
times than are feasible from first principles, but unfortunately no
existing potential is capable of quenching directly from the liquid to
the amorphous phase. Instead, empirical model liquids typically
transform into glassy phases upon cooling, characterized by frozen-in
liquid structure
\cite{broughton,luedtke,kluge,ding,biswas}. Alternatively, in the case
of the T2 potential, quenching results in a reasonable amorphous
structure \cite{t2,kelires}, but the original liquid phase is not
realistic and already contains excessive tetraherdal order \cite{t2}.
Therefore, it has been impossible to simulate an experimentally
relevant path to the amorphous structure ({\it e.g.}  laser quenching
\cite{roorda}), and artificial preparation methods have been required
to create large-scale amorphous structures with empirical potentials
\cite{broughton,luedtke,kluge,ding,www,hm}. For example, the so-called
indirect SW amorphous (ISW) structure is created by increasing the
strength of the SW three-body interaction during the quench
\cite{broughton,luedtke,kluge}. The ISW structure is stable with the
unaltered SW potential, but since it has 81\% four-fold coordinated
atoms \cite{broughton}, it only bears a weak resemblance to the real
amorphous structure \cite{ding}. Abandoning molecular dynamics, an
improved amorphous phase with close to 87\% four-fold coordination can
be generated using the bond-switching algorithm of Wooten, Winer and
Weaire \cite{www}, but such a method does not permit accurate
simulation of atomic motion. A more realistic, large-scale amorphous
structure can be prepared using a hybrid of these methods: Holender
and Morgan create an amorphous structure of over $10^5$ atoms with
almost 94\% four-fold coordination by patching together a number of
smaller WWW structures, thermally treating the sample at high
temperatures and then relaxing it using molecular dynamics with the SW
potential modified for stronger angular forces \cite{hm}. As these
authors emphasize, however, this preparation method (which we call the
HM2 model) was designed by trial-and-error to fit the experimental
structure factor and bears no resemblance to the real experimental
generation of a-Si. They also report that if the SW potential is not
modified, their preparation method results in a structure (which we
call the HM1 model) with only 74\% four-fold coordination.

Remarkably, the present model predicts a quench directly from the
liquid into a high-quality amorphous structure. The phase transition
is quite robust, since it occurs even with fast cooling rates. For
example, quenching at -300 K/ps leads to a reasonable structure with
84\% four-fold coordination. At much slower quench rates of -1 K/ps,
an improved structure of 1728 atoms at $T = 300$ K and zero
pressure is produced with almost 95\% four-fold coordination.  The
excess enthalpy of the amorphous phase compared with the crystal is
0.22 eV/atom, closer to experimental values $\leq 0.19$ eV/atom
\cite{roorda} than the {\it ab initio} value 0.28 eV/atom (probably
due to the constrained volume and small system size used in the {\it
ab initio} study \cite{stichamo}).

The coordination statistics of the amorphous phase obtained with our
model, given in Table \ref{tabamo}, are closer to {\it ab initio}
results \cite{stichamo} than with most of the empirical models
described above. The HM2 model provides a comparable description, but
we stress that its preparation procedure is unphysical and that the
modification of the SW potential necessary to achieve the improved
description (see the difference between HM1 and HM2 in Table
\ref{tabamo}) degrades many important properties, such as elastic
constants, defect formation energies and the melting point. Since
realistic preparation methods and dynamics have not been achieved with
interatomic potentials, in the following we compare our results only
with experiments and {\it ab initio} simulations.

\begin{figure}[h]
\begin{center}
\leavevmode
\mbox{
\psfig{figure=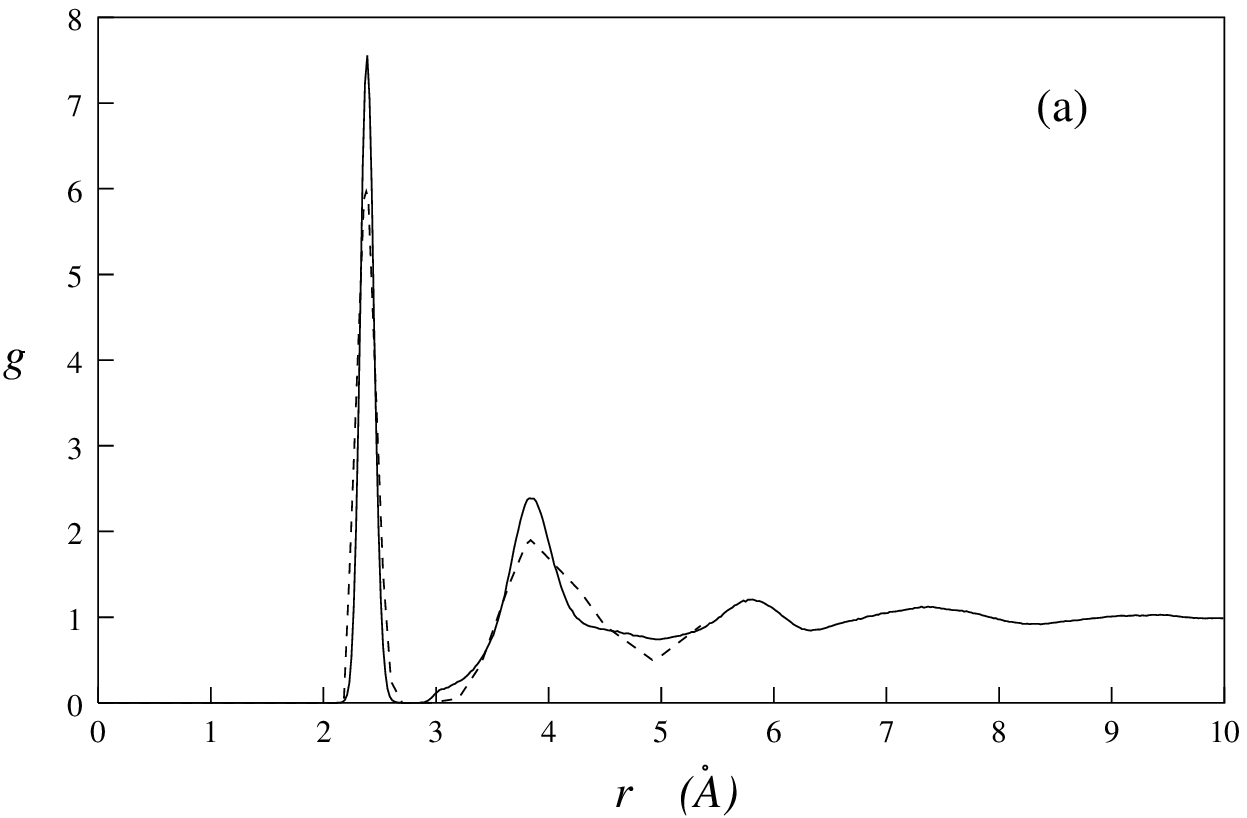,width=4in}
}
\mbox{
\psfig{figure=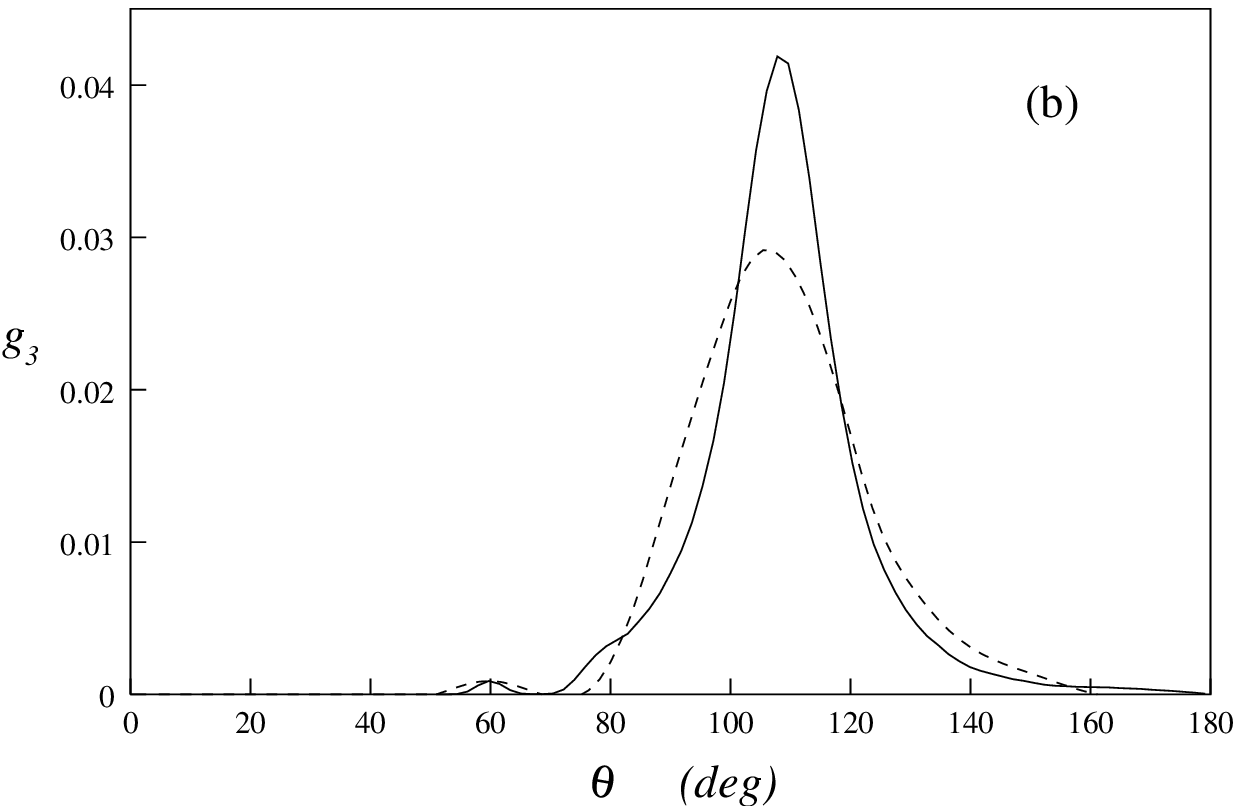,width=4in}
}
\begin{minipage}[h]{5in}
\caption{(a) Pair correlation function and (b) bond angle distribution
for the amorphous phase at $T = 300$ K and zero pressure with the
present potential (solid lines) and the {\it ab initio} model of
Ref. \protect{\cite{stichamo}} (dashed lines).  }
\label{figamo}
\end{minipage}
\end{center}
\end{figure}

\begin{figure}[h]
\begin{center}
\leavevmode
\mbox{
\psfig{figure=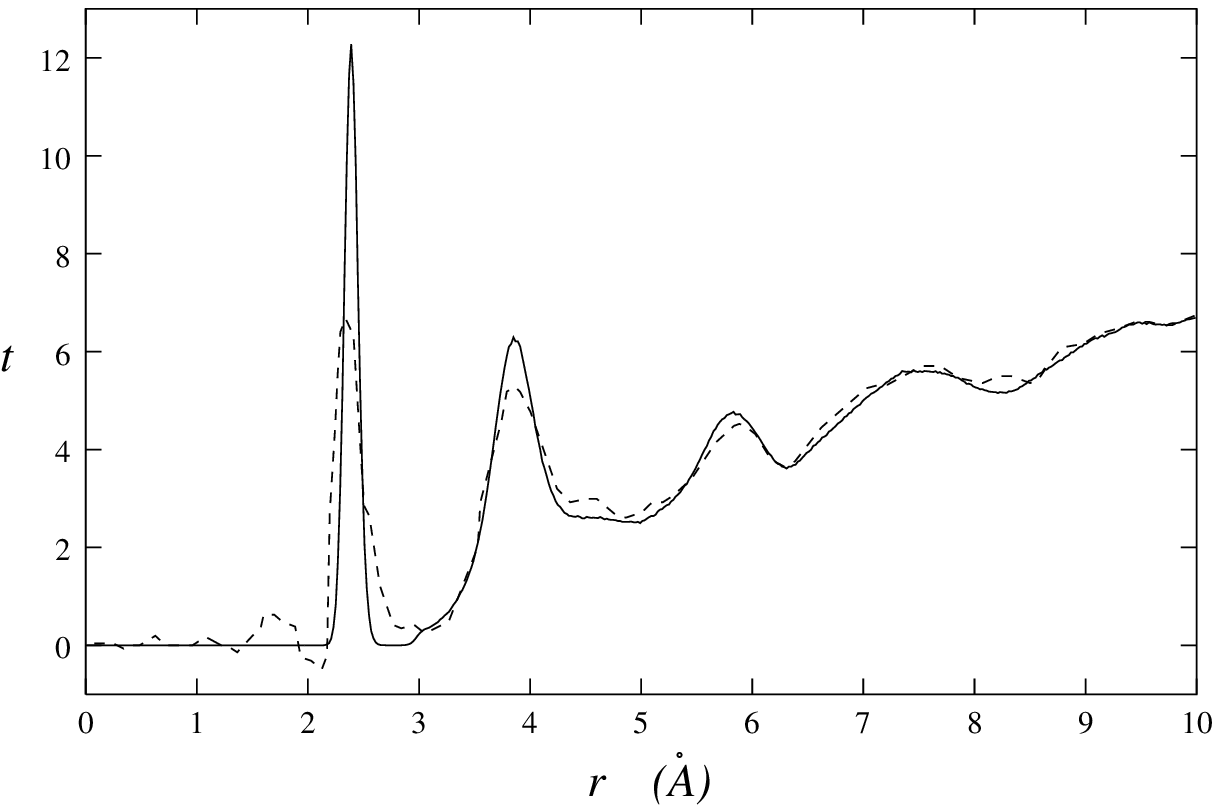,width=4in}
}
\begin{minipage}[h]{5in}
\caption{Radial distribution function $t(r) = 4\pi\rho r g(r)$ for the
amorphous phase at room temperature and zero pressure using our model,
compared with the results of neutron scattering experiments on
pure evaporated-beam-deposited a-Si thin films by Kugler {\it et. al.}
\protect\cite{kugler}.
}
\label{figamoexpt}
\end{minipage}
\end{center}
\end{figure}

The pair correlation function shown in Fig. \ref{figamo} (a) is in
good agreement with {\it ab initio} results \cite{stichamo}. Moreover,
Fig.  \ref{figamoexpt} shows that the radial distribution function
$t(r) = 4 \pi \rho r g(r)$ is in excellent agreement with the results
of neutron scattering experiments by Kugler {\it et. al.}
\cite{kugler} (using their experimental density $\rho=0.054$
atoms/\AA$^3$ for comparison).  The persistence of intermediate-range
order up to 10 \AA captured by our model as in experiment is a
strength of the empirical approach, since this distance is roughly the
size of the periodic simulation box used in the {\it ab initio}
studies \cite{stichamo}.  Given the limited resolution of the
experimental data, especially at small $r$ (large $q$ in the structure
factor), the sharper first three peaks with our model may be
interpreted as refinements of the experimental results. In Table
\ref{tabamoexpt} we summarize a detailed comparison of features of
a-Si as obtained with our model and from ab-initio results, against
those revealed by experiment.  Overall the agreement between
experiment is very satisfactory, with the results of the present model
somewhat closer to experimental values than {\it ab initio} results 
as in the case of the enthalpy ($\Delta H_{a-c}$) and the bond-length 
($\sigma_{r_1}$) and bond-angle ($\sigma_\theta$) deviations.

The bond angle distribution $g_3(\theta,r_m)$ shown in
Fig. \ref{figamo} (b) is narrowly peaked just below the tetrahedral
angle, and also reproduces the small, well-separated peak at
60$^\circ$ observed in {\it ab initio} simulations \cite{stichamo}
(unlike in previous empirical models).  The average angle is
108.7$^\circ$ and standard deviation 13.6$^\circ$ in close agreement
with the experimental values\cite{fortner} of 108.6 $\pm 0.2^\circ$
and 10.2 $\pm 0.8^\circ$ and {\it ab initio} values\cite{stichamo} of
108.3$^\circ$ and 15.5$^\circ$, respectively. Notice that the peaks in
both $g(r)$ and $g_3(\theta,r_m)$ are narrower and taller with our
model than with {\it ab initio} methods, which probably reflects the
small system size and short times of the {\it ab initio} simulations
compared to ours.  In summary, our potential reproduces the random
tetrahedral network of amorphous silicon very well, following a
realistic preparation procedure that starts with a liquid phase and
cools it down without any artificial changes.

\section{Conclusion}

We have developed a new empirical potential for silicon that provides
a considerable improvement over existing models in describing local
structures away from equilibrium.  The model introduces a
theoretically motivated functional form that incorporates several
coordination dependent functions to adapt the interactions for
different coordinations. The fitted potential faithfully reproduces
the elastic constants of the equilibrium DC structure and also
captures the energetics of a wide range of point and extended
defects and related activation energies. The superior description of
bulk phases and defects is achieved with only thirteen fitting
parameters, indicating exceptional transferability of the EDIP functional
form. Its extended range of success over existing models with
comparable numbers of parameters cannot be attributed to fitting alone.
 
For dislocations, this is the first empirical model to give a full
description of core properties. It predicts the correct reconstruction
for both the $90^\circ$- and $30^\circ$-partial dislocations, and the
reconstruction energies are in agreement with {\it ab initio} data.
The bond stretching  is also in good agreement
with {\it ab initio} results, pointing to the fact that this model
predicts reasonably accurately not only the energies but also the
local structure of dislocation cores.

This is the also the first empirical model to predict a quench
directly from the liquid to the amorphous phase. The quality of the
resulting amorphous phase, with almost 95\% four-fold coordination, is
better than with any existing empirical preparation method. In some
ways the amorphous phase with our model is even somewhat closer to
experiment than with {\it ab initio} simulations (surely because the
latter are limited to very small system sizes and very short times),
which is an encouraging success of the empirical approach to materials
modeling.

Our model possesses the same level of efficiency as the SW potential,
and simulations involving thousands of atoms may be readily performed
on typical workstations.  Therefore, it holds promise for successful
applications to several systems that are still inaccessible to {\it ab
initio} calculations and are outside the range of validity of other
empirical models.  Taking into account the success of our model in
describing dislocation properties, we suggest that it may also provide
a reasonable description of small-angle grain boundaries and other
such extended bulk defects. Considering its success with the amorphous
and crystalline phases, the model may also describe the
a-c interface and solid-phase epitaxial growth.

\acknowledgments 

Partial support was provided to JFJ, VVB, and SY by the MRSEC Program
of the National Science Foundation under award number DMR 94-00334.
Partial support was provided to MZB by a Computational Science
Graduate Fellowship from the Office of Scientific Computing of the
US Department of Energy and by the Harvard MRSEC , which is funded
by NSF grant number DMR-94-00396.  JFJ also acknowledges partial
support from Brazilian Agency FAPESP.

\end{multicols}

\begin{center}
\begin{table}
\label{table1}
\begin{minipage}[t]{9.5cm}
\caption{Values of the parameters that define the potential, obtained
from a simulated annealing fit to the database described in the text.}
\label{tab1}
\begin{tabular}{ccc}
 $A$ = 7.9821730   eV & $B$ =  1.5075463    \AA & $\rho$ = 1.2085196  \\
 $a$ = 3.1213820  \AA & $c$ = 2.5609104  \AA & $\sigma$ = 0.5774108 \AA \\
$\lambda$ =  1.4533108 eV & $\gamma$ = 1.1247945 \AA & $\eta$ = 0.2523244 \\ 
$Q_0$ = 312.1341346 & $\mu$ = 0.6966326  & $\beta$ = 0.0070975   \\ 
$\alpha$ = 3.1083847 & & \\
\end{tabular}
\end{minipage}
\end{table}
\end{center}

\begin{center}
\begin{table}
\begin{minipage}[t]{9.5cm}
\caption{Energy and lattice parameters for high symmetry structures.
Here we consider the ground-state diamond cubic (DC), face centered
cubic (FCC), body centered cubic (BCC), simple cubic (SC)
and hexagonal close-packed (HCP) crystals.  For DC,
the cohesive energy per atom $E_c^{DC}$ is given in eV, while for the
other crystals the difference of the cohesive energy $E_c$ from the
ground state DC crystal, $\Delta E = E_c - E_c^{DC}$, is given.  All
lattice constants $a_o$ are for the conventional unit cells in {\AA}.
For the hexagonal crystals we also give the $c/a$ ratios. We also
compute the lattice constant and binding energy of an isolated
hexagonal plane (HEX). For this comparison we use the SW potential
with the rescaled cohesive energy for the ground state, as described
in Ref. \protect{\cite{bala}}.}
\label{tab2}
\begin{tabular}{lcccccc} 
      &        & DFT/LDA &  EDIP &   SW   &   T2  & T3    \\ 
\hline 
   DC  & $E$   & -4.65   & -4.650 & -4.63 & -4.63  & -4.63 \\
     & $a_o$   &\ 5.43   &  5.430 & 5.431 & 5.431  & 5.432 \\
\hline
SC &$\Delta E$ &\ 0.348  &\ 0.532 &\ 0.293 &\ 0.343 &\ 0.318 \\
 & $a_o$       &\ 2.528  &\ 2.503 &\ 2.612 &\ 2.501 &\ 2.544 \\ 
\hline
BCC &$\Delta E$&\ 0.525  &\ 1.594 &\ 0.300 &\ 0.644 &\ 0.432 \\ 
    & $a_o$    &\ 3.088  &\ 3.243 &\ 3.245 &\ 3.126 &\ 3.084 \\ 
\hline
FCC &$\Delta E$&\ 0.566  &\ 1.840 &\ 0.423 &\ 0.548 &\ 0.761 \\
    & $a_o$    &\ 3.885  &\ 4.081 &\ 4.147 &\ 3.861 &\ 3.897 \\ 
\hline
HCP &$\Delta E$&\ 0.552  &\ 0.933 &\ 0.321 &\ 0.551 &\ 0.761 \\ 
    & $a_o$    &\ 2.735  &\ 2.564 &\ 3.647 &\ 2.730 &\ 2.756 \\
    & $c/a$    &\ 1.633  &\ 2.130 &\ 0.884 &\ 1.633 &\ 1.633 \\ 
\hline
HEX &$\Delta E$&\ 0.774  &\ 0.640 &\ 1.268 &\       &\  \\
    & $a_o$    &\ 3.861  &\ 4.018 &\ 4.104 &\       &\ 
\end{tabular}
\end{minipage}
\end{table}
\end{center}

\begin{center}
\begin{table}
\begin{minipage}[t]{9.5cm}
\caption{Elastic constants $C_{11}, C_{12}, C_{44}, C_{44}^\circ$ (without
internal relaxation) and bulk modulus $B$ of the diamond cubic
structure in Mbar, and the values of two combinations, $C_{11}-C_{12}$
and $C_{12}-C_{44}$, that are important for stability. The
experimental values are from Ref. \protect{\cite{simm}} and the
tight-binding results from Ref. \protect{\cite{noam}}. The {\it ab
initio} result for $C_{44}^\circ$ (LDA) is from Ref. \protect\cite{nielsen}. }
\label{tab3}
\begin{tabular}{cccccccc} 
                & EXPT  &   LDA & EDIP  &   SW &  T2  &  T3  & TB  \\
\hline 
$C_{11}$        &\ 1.67 &       &\ 1.75 & 1.61 & 1.27 & 1.43 & 1.45\\ 
$C_{12}$        &\ 0.65 &       &\ 0.62 & 0.82 & 0.86 & 0.75 & 0.85\\ 
$C_{44}$        &\ 0.81 &       &\ 0.71 & 0.60 & 0.10 & 0.69 & 0.53\\ 
$C_{44}^\circ$  &       &\ 1.11 &\ 1.12 & 1.17 & 0.92 & 1.19 & 1.35\\ 
$B$             &\ 0.99 &       &\ 0.99 & 1.08 & 0.98 & 0.98 & 1.05\\ 
$C_{11}-C_{12}$ &\ 1.02 &       &\ 1.13 & 0.79 & 0.41 & 0.68 & 0.60\\
$C_{12}-C_{44}$ & -0.16 &       & -0.09 & 0.22 & 0.76 & 0.06 & 0.32\\
\end{tabular}
\end{minipage}
\end{table}
\end{center}

\begin{center}
\begin{table}
\begin{minipage}[t]{9.5cm}
\caption{Ideal (unrelaxed) formation energies $E_f^{ideal}$ of point
defects (in eV) and relaxation energies $\Delta E = E_f^{ideal} -
E_f^{relaxed}$ using a 54 atom unit cell.  The {\it ab initio}
(DFT/LDA) results are from Refs.
\protect{\cite{kaxlda,pand,baryam,kelly,seong}} and tight-binding
results from Ref. \protect{\cite{noam}}. }
\label{tab4}
\begin{tabular}{lccccccc} 
     &            & DFT/LDA & EDIP &  SW   & T2   &  T3  & TB \\  \hline
$V$  & $E_f$      & 3.3-4.3 & 3.47 &\ 4.63 & 2.83 & 4.10 & 4.4 \\ 
     &$\Delta E_f$& 0.4-0.6 & 0.25 &\ 1.81 & 0.02 & 0.40 & 1.2 \\ \hline
$I_T$& $E_f$      & 3.7-4.8 & 6.15 & 12.21 & 5.85 & 6.92 & 4.5 \\
     &$\Delta E_f$& 0.1-0.2 & 2.10 &\ 6.96 & 0.82 & 3.47 & 0.5 \\ \hline
$I_H$& $E_f$      & 4.3-5.0 & 6.86 & 17.10 & 5.39 & 8.22 & 6.3 \\
     &$\Delta E_f$& 0.6-1.1 & 2.70 & 10.15 & 1.72 & 3.61 & 1.3 \\ \hline
$CE$ & $E_f$      &   5.47  & 5.41 &\ 7.90 & 6.50 &      & 5.5 \\
     &$\Delta E_f$&   0.90  & 0.59 &\ 3.26 &      &      & 1.8 \\
\end{tabular}
\end{minipage}
\end{table}
\end{center}

\begin{center}
\begin{table}
\begin{minipage}[t]{9.5cm}
\caption{Unstable Stacking Fault energy (in $J/m^2$) for the unrelaxed
glide and shuffle $\{111 \}$ planes as obtained from calculations
using DFT/LDA \protect{\cite{kax1}}, SW, and the present interatomic
potential.}
\label{tab4a}
\begin{tabular}{ccccccc} 
        &         & DFT/LDA & EDIP  &  SW   \\  \hline
glide   & $<112>$ &\  2.51  &\ 3.24 &\ 4.78 \\
        & $<110>$ &  24.71  & 13.45 & 26.17 \\ \hline
shuffle & $<110>$ &\  1.84  &\ 2.16 &\ 1.38 \\
\end{tabular}
\end{minipage}
\end{table}
\end{center}

\begin{center}
\begin{table}
\begin{minipage}[t]{9.5cm}
\caption{Reconstruction energy (in eV/$b$) and APD energy (in eV) for
core structures of partial dislocations, where $b$ is the repeat
distance of the dislocation. The DFT/LDA result for reconstruction of
the 90$^\circ$-partial dislocation is from Ref. \protect{\cite{bigg}}, and
for the 30$^\circ$-partial dislocation is from
Ref.\protect{\cite{arias2}}.  Results for the interatomic potential
calculations (SW and T2) are from Ref. \protect{\cite{duesb1}}.
Tight-binding result for the 90$^\circ$-partial dislocation is from
Ref. \protect{\cite{nunes}}.}
\label{tab5}
\begin{tabular}{cccccc} 
                  & DFT/LDA & EDIP & SW & T2 & TB \\ \hline
 Reconstruction & & & & & \vspace{0.05cm} \\ 
90$^\circ$-partial & 0.87 & 0.84 & - &\ 0.37 & 0.68 \\ 
30$^\circ$-partial & 0.43 & 0.33 & 0.81 & -0.13 & \\ \hline 
\vspace{0.05cm} APD & & & & \\
 90$^\circ$-partial & - & 0.41 & - &\ 0.37 & 1.31 \\ 
30$^\circ$-partial & 0.43 & 0.34 & 0.84 & -0.13 & \\
\end{tabular}
\end{minipage}
\end{table}
\end{center}

\begin{table}
\begin{center}
\begin{minipage}[t]{9.5cm}
\begin{center}
\begin{tabular}{ccccccc}
 & \multicolumn{2}{c}{Direct Quench} & \multicolumn{4}{c}{Artificial
Preparation} \\
       & LDA   &  EDIP   &  HM2  & WWW  &  ISW  &  HM1 \\
\hline 
 $N_3$ & \ 0.2 & \ 0.23  &\ 2.18 &\ 1.2 & \ 0.0 &\ 0.0 \\
 $N_4$ &  96.6 &  94.43  & 93.74 & 86.6 &  81.0 & 73.6 \\
 $N_5$ & \ 3.2 & \ 5.34  &\ 4.04 & 11.8 &  18.1 & 24.6 \\
 $N_6$ & \ 0.0 & \ 0.00  &\ 0.01 &\ 0.2 & \ 0.9 &\ 1.5 \\ 
\end{tabular}
\end{center}
\caption{ Coordination statistics for model amorphous structures at
room temperature. $N_n$ is the percentage of atoms with $n$ neighbors
closer than the first minimum of $g(r)$.  We compare structures
generated by molecular dynamics simulation of a direct quench from the
liquid using an {\it ab initio} method\protect\cite{stichamo} (LDA) and our
interatomic potential (EDIP) with structures generated by various
artifical preparation methods described in the text: HM2\protect\cite{hm},
WWW\protect\cite{www}, ISW\protect\cite{broughton} and HM1\protect\cite{hm}.}
\label{tabamo}
\end{minipage}
\end{center}
\end{table}

\begin{table}
\begin{center}
\begin{minipage}[t]{6.5in}
\begin{center}
\begin{tabular}{lccccccccc}
 & $\rho_a$ & $\Delta H_{a-c}$ & $\overline{Z}$ & $\overline{r}_1$ & $\sigma_{r_1}$
        & $\overline{r}_2$ & $\overline{r}_3$ & $\overline{\theta}$ & $\sigma_\theta$ \\
\hline
EDIP & 0.04836      & 0.22     & 4.054      & 2.39       & 0.034      
        & 3.84 & 5.83 & 108.6 & 14.0 \\
EXPT & 0.044--0.054 & $<$ 0.19 & 3.90--3.97 & 2.34--2.36 & 0.07--0.11 
        & 3.84 & 5.86 & 108.6 & 9.4--11.0 \\
LDA  &      -       &  0.28    & 4.03       & 2.38       & 0.079      
        & 3.84 &  -   & 108.3 & 15.5
\end{tabular}
\end{center}
\caption{Comparison of thermodynamic and structural properties of the
present model (EDIP) for a-Si with (annealed) {\it ab initio}\cite{stichamo} 
and with (annealed) experimental\cite{stichamo,kugler,fortner,roorda} results.
Shown  are the density $\rho_a$ in \AA$^{-3}$, the excess enthalpy $\Delta
H_{a-c}$ compared to the crystal in eV/atom, the average coordination
$\overline{Z}$, the mean $\overline{r}_1$ and  standard deviation 
$\sigma_{r_1}$ of the first neighbor distance in \AA, the mean 
second $\overline{r}_2$ and third $\overline{r}_3$ neighbor distances 
in \AA and the mean $\overline{\theta}$ and standard deviation 
$\sigma_\theta$ of the first neighbor bond angles in degrees. }
\label{tabamoexpt}
\end{minipage}
\end{center}
\end{table}

\vspace*{2.cm}

\begin{references}
\bibitem{kaxlda} The fitting database consists of DFT/LDA total
energies computed using a plane wave basis with at least a 12 Ry
cutoff and adequate sampling of the Brillouin zone for reciprocal
space integrations. Supercells for the point defect calculations
included 53-55 atoms, so that long range elastic relaxation energies,
on the order of 0.01 eV, are ignored. Stacking fault calculations
involved 12 atom supercells. For example, the formation energies for
the unrelaxed defects are: 3.3 eV for the vacancy, 3.7 eV and 4.3 eV
for the tetrahedral and hexagonal interstitials, and 5.47 eV for the
concerted exchange saddle point configuration.
\bibitem{ishinote} A bimodal bond angle disribution, similar to the
present one, has been reported for the T3 potential \cite{ishi}, but
the authors found it necessary to arbitrarily extend the coordination
cutoff beyond the first minimum of $g(r)$ to obtain such
results. These simulations were also performed at very high
temperature (3000 K) and constant volume (set to the volume of the
experimental liquid, not the T3 liquid).
\bibitem{cutoff_note} Similar conclusions have been reached regarding
the short cutoff of the Tersoff potentials \cite{t2,t3,ishi}.
\end{references}
\end{document}